# Physical principles underpinning molecular-level protein evolution

Jorge A. Vila

IMASL-CONICET, Universidad Nacional de San Luis, Ejército de Los Andes 950, 5700 San Luis, Argentina.

Since protein mutations are the main driving force of evolution at the molecular level, a proper analysis of them—and the factors controlling them—will enable us to find a response to several crucial queries in evolutionary biology. Among them, we highlight the following: At the molecular level, what factors determine whether protein evolution is repeatable? Aiming at finding an answer to this and several other significant questions behind protein evolvability and the factors that control it—including, but not limited to, the proteins' robustness, the evolutionary pathways, the number of ancestors, the epistasis, the post-translational modifications, and the location and order of mutations—we distinguish two evolutionary models in our analysis: convergent and divergent, based on whether or not a "target sequence" needs to be reached after $n$-mutational steps beginning with a wild-type protein sequence (from an unknown ancestor). Preliminary results suggest—regardless of whether the evolution is convergent or divergent—a tight relationship between the thermodynamic hypothesis (or Anfinsen's dogma) and protein evolution at the molecular level. This conjecture will allow us to uncover how fundamental physical principles guide protein evolution, and to gain a deeper grasp of mutationally driven evolutionary processes and the factors that influence them. Breaking down complex evolutionary problems into manageable pieces—without compromising the vision of the problem as a whole—could lead to effective solutions to critical evolutionary biology challenges, paving the way for further progress in this field.

**Introduction**

Let us start by thinking about the Conformal Cyclic Cosmology theory according to which "…*the big bang was not actually the origin of our universe, but the continuation of the remote future of a previous aeon. So the universe expands and contracts and then indulges in this exponential expansion which we now see in our own aeon, where the expansion of the universe accelerates. And it continues*…" (Penrose, 2022). Even though we are not interested in discussing the validity of the theory or its foundations, such a model of the universe opens an immense number of interrogations in our minds, one of which is a key question from an evolutionary biology perspective: *how* likely is life to occur in some of the cycles—aeon—of the universe? If it does indeed happen, would the evolutionary pathways lead to a very different ending? Over time, many scholars have become interested in finding an answer to such an important question (Gould, 1989; Lobkovsky & Koonin, 2012; Powell, 2012; Dobzhansky, 2013; Orgogozo, 2015; Blount *et al.*, 2018; Yamasaki & Kitano, 2021; Wortel *et al.*, 2022; Nosil *et al.*, 2024; Pearless & Freed, 2024). Rather than addressing a challenge of this magnitude, we have begun with more basic issues, such as investigating fundamental evolutionary events—at the molecular level—to better understand how life on Earth could have evolved over billions of years. Within this outline, we focus on analyzing protein evolution due to mutations, aiming at comprehending and deciphering the fundamental physical principles underpinning the molecular-level evolutionary processes of proteins and determining to what extent they could be repeatable. Answering such challenges accurately requires addressing several questions, such as: To what extent does the validity of the thermodynamic hypothesis (or Anfinsen's dogma; Anfinsen, 1973) impact on protein evolvability? Will different evolutionary pathways be equally probable? If they are not, what factors dictate the evolutionary preference among them? What effect do mutations' location and order have on the evolutionary process? Could we reverse the evolution of proteins at the molecular level to their starting sequence? Considering that a mutation could either be an amino acid substitution or a post-translational modification (Vila, 2024a), how will the above analysis vary depending on the type of mutation occurring? In any of the above situations, what will the role and impact of epistasis be? For example, if there are changes in the milieu (pH, temperature, solvent, ionic strength, *etc.*), how will they impact on the outcomes of protein evolution at the molecular level? It is worth noting that the term "outcome" of the evolutionary process refers here to a functional protein with a unique amino-acid sequence characterized by observables parameters such as marginal stability (Vila, 2021) or, equivalently, folding rate (Vila, 2023b), amide hydrogen-exchange protection factors (Vila, 2022), or catalytic activity (Vanella *et al*., 2024).

Before proceeding, five key concepts for the analysis must be either remembered, defined, or clarified—specifically, the epistasis and its origin (molecular basis); the notions of reversible and repeatable protein evolution; the idea of "target sequence" as a model of natural selection; the thermodynamic hypothesis, also known as Anfinsen's dogma (Anfinsen, 1973); and, finally, the origin of a free-energy threshold for the maximum change of protein stability allowed.

Firstly, it is common knowledge that epistasis results from either the genetic context changes in which mutations occur, or the combined effect of two or more mutations deviating from that predicted result by adding their individual effects (LiCata & Ackers, 1995; Phillips, 1998; Grant & Grant, 2002; Cordell, 2002; Weinreich *et al*., 2005; Weinreich *et al*., 2006; Ortlund *et al*., 2007; Phillips, 2008; de Visser *et al*, 2011; Breen *et al*, 2012; McCandlish *et al*., 2013; Ashenberg *et al*., 2013; Orgogozo, 2015; Starr & Thornton, 2016; Miton & Tokuriki, 2016; Sailer & Harms, 2017a; Sailer & Harms, 2017b; Adams *et al*, 2019; Domingo *et al*., 2019; Miton *et al*., 2020; Miton *et al*., 2021; Park *et al*., 2022; Jayaraman *et al*, 2022; Buda *et al*., 2023; Diaz *et al*., 2023). Not as widely known, though, is the relationship between the origin (molecular basis) of mutational epistasis and the protein folding problem (Vila, 2024b)—one of the yet-unsolved biggest challenges of structural biology—whose solution demands a precise understanding of *how* a sequence encodes its folding. Unfortunately, tackling the latter requires solving an *n*-body problem (Vila, 2023a). The fact that even the effect of a single mutation cannot be accurately predicted (Zheng *et al*., 2024) only exacerbates the accurate foreseeing of epistasis effects. Secondly, the notion of reversible and repeatable protein evolution process refers—within the framework of this study—to the ability of a protein to either revert to its original state (the wild-type protein sequence) or evolve predictably (toward a given target sequence). Understanding these concepts is critical because they should provide useful insights into the mechanisms beneath protein evolution and adaptation. A reversible protein evolution process, for example, could occur when a series of mutations that result in undesirable biological changes, such as function loss, are corrected by subsequent mutations to restore to their original sequence. In contrast, a repeatable protein evolution process involves mutations occurring independently in different evolutionary pathways to achieve the same functional form (target sequence). Thirdly, the concept of "target sequence" arises as a model of natural selection (Maynard Smith, 1970), in which one starting

sequence—written using the one-letter code representation of the naturally occurring amino acids—can be converted, under certain rules, into a chosen sequence by changing (or mutating) one letter at a time (Ogbunugafor, 2020). Fourthly, it becomes essential to define and understand the thermodynamic hypothesis (or Anfinsen's dogma), which states that—for a given amino-acid sequence and milieu—the protein native state is the conformation with the lowest accessible Gibbs free energy, or, to put it more precisely, the one for which "…*the Gibbs functional G($\mathbb{Z}\{\phi,\psi,\chi\}$)$_{milieu}$ has the lowest free-energy minimum with respect to all possible distributions of $\mathbb{Z}\{\phi,\psi,\chi\}$, where ($\phi,\psi,\chi$) are the protein torsional angles…*" (Vila, 2020). Finally, the term "protein marginal stability" is rooted in the Gibbs free-energy difference between the native and first unfolded states (Hormoz, 2013; Vila, 2019; Vila, 2021). The existence of a threshold to the allowed Gibbs free energy variations (~7.4 Kcal/mol) arises as a consequence of the thermodynamic hypothesis validity—and not as a product of evolution (Martin & Vila, 2020)—as proved by using a statistical mechanics analysis (Vila, 2019). We have also presented sound evidence that such a threshold is of universal validity, *i.e.*, it stands for proteins of any fold class, sequence, or protein size. Last but not least, such a threshold also provides a physical substrate for the neutral (Kimura, 1968) or nearly neutral (Ohta, 1973) theory of evolution to occur (Martin & Vila, 2020). Indeed, if mutations cause marginal stability changes that exceed the free energy threshold, the protein will unfold or become nonfunctional. Therefore, the remaining mutations could only moderately destabilize or stabilize a protein structure.

Overall, within the above set-in conditions and definitions, we will analyze a simple protein evolution model that—beginning with a wild-type protein sequence (from an unknown ancestor)—distinguishes between two types of evolutionary pathways depending on where they end after *n*-consecutive mutational steps. The model will be referred to as "convergent protein evolution" if all pathways result in the same target sequence (see Figure 1a-c). Otherwise, they will be referred to as "divergent protein evolution" (see Figure 2a). Lastly, the evolution of proteins will be examined via a molecular approach.

## Analysis of Two Protein Evolution Models
## I.- Convergent protein evolution

In the presence of a target sequence, we can quickly determine whether or not this evolutionary model is repeatable—and perhaps reversible—by providing answers to the questions below.

### 1.- How do different pathways/trajectories affect a convergent evolutionary process?

The existence of different mutational pathways (shown in Figure 1a)—characterized by the location, order, and type of mutation, *i.e.*, amino acid substitutions or post-translational modification—does not affect the outcomes of convergent evolution, *e.g.*, in terms of the protein folding rates or amide hydrogen exchange protection factors (Vila, 2024a). The latter is a consequence of the fact that starting with a wild-type protein (*wt*), the total change in the Gibbs free energy after *n*-mutational steps to reach the target sequence (*ts*) ($\Delta\Delta G_n = \Delta G_{ts} - \Delta G_{wt}$, $\forall$ $n \geq$ 1) is—from a thermodynamic perspective—a state function and, thus, independent of the pathways followed by the evolutionary process (Vila, 2022). According to this viewpoint, a convergent protein evolution can be viewed as a black box (see Figure 1b), in which the native-state-protein-marginal stability for both the starting wild-type sequence ($\Delta G_{wt}$) and the end target sequence ($\Delta G_{ts}$) is the only information that is needed to analyze the evolutionary process; here $\Delta G_x$ (with *x* = *wt* and *ts*) represents the Gibbs free energy gap between the native state of protein *x* and the conformation at the highest point of the free-energy profile (Vila, 2021)—beyond which the protein unfolds or becomes nonfunctional (Vila, 2023b).

It is important to be cautious when overlooking the impact of different pathways for an evolutionary convergent process. Indeed, consideration of the speed of evolution—which transforms pathways into trajectories—will be crucial in the analysis because, as previously shown (Vila, 2024a), the total time needed to evolve from the wild-type sequence (*wt*) to the target sequence (*ts*) will vary among all potential evolutive trajectories since the location (but not the order) of mutations affects the marginal stabilities of proteins, and, hence, their folding rates (Vila, 2023b). The latter implies that many evolutionary trajectories may not exist, not just due to unfavorable intramolecular interactions (Weinreich *et al*., 2006), but also because they may require prohibitive biological time to reach a target sequence.

## 2.- Is a model of convergent protein evolution always repeatable?

From a statistical-thermodynamic perspective (Hill, 1960), the repeatability of a convergent evolutionary process after $n$ mutational steps is not binary but determined by the relative probability ($\Omega$) of reaching the target sequence (*ts*) with regard to that of the wild-type (*wt*) sequence, as given by $\Omega = P_{ts}/P_{wt} = e^{\beta \Delta \Delta G_n}$; where $\beta = 1/RT$, *with R being the gas constant* and $T$ the absolute temperature in ºK, $P_x = e^{\beta \Delta G_x}/Q$ the probability of a given sequence $x$, Q = $\sum_{\{\mu \ni \xi\}} e^{\beta \Delta G_\mu}$ the Partition Function (Hill, 1960), $\Delta G_\mu$ the Gibbs free energy gap between the native state of protein $\mu$ and a native-like-conformation at the highest point of the free-energy profile (Vila, 2019; Vila, 2021), and $\mu$ an index running over the whole protein sequence space ($\xi$) that contains different functional proteins between ~$10^{16}$ and ~$10^{19}$, regardless of whether the mutations include posttranslational modifications or not (Vila, 2024b). Then, we can distinguish between two possible scenarios for a convergent evolutionary process. Firstly, let us assume that $\Delta \Delta G_{\overrightarrow{n}} > 0$; where the arrow ($\rightarrow$) denotes the direction of evolutionary pathways, as seen in Figure 1b, starting with the wild-type protein (*wt*) and ending with the target sequence (*ts*). In this case, the convergent evolutionary process will probably be repeatable because $\Omega > 1$, although not reversible ($\Omega < 1$) because their occurrence (reversibility) will fall off exponentially. Secondly, let us assume $\Delta \Delta G_{\overrightarrow{n}} < 0$, as shown in Figure 1c. In this alternative case, the convergent evolutionary process will probably be reversible ($\Omega > 1$), albeit not repeatable ($\Omega < 1$). These scenarios have shown, firstly, that convergent evolution does not guarantee that the outcome of a process will always be repeatable and, secondly, that nature may repair—from a thermodynamic perspective—undesired evolutionary changes. For example, a species may develop a new trait that proves detrimental in a short time. Still, through reverse evolution (Teotónio & Rose, 2001), it may be possible for the species to revert to a previous state where the trait was not present.

## 3.- What happens if more than one ancestor is present?

The term "convergent" formerly referred to a protein evolution model that began from a single common ancestor and, as such, should be redefined to be consistent with the accepted terms in the literature in light of protein evolution models that begin from multiple ancestors. Consequently, when proteins evolve from $j$-ancestors (similar or distinct, with $j > 1$) to the same—or similar—target sequence (Bolnick *et al.*, 2018), as shown in Figure 1d, the processes could be referred to as "parallel" or "convergent" because their distinction lies on the ancestor origin. To put it simply, if the protein evolves from similar ancestors, it is named "parallel," whereas if it evolves from distinct ancestors, it is named "convergent" (Cerca, 2023). In any case, the evolutionary process from each $j$-ancestor should be treated as described in Section I.2, and, hence, those leading to a $\Delta\Delta G_j^{\rightarrow} > 0$ will be likely repeatable ($\Omega > 1$), albeit unlikely irreversible, while the contrary will occur if $\Delta\Delta G_j^{\rightarrow} < 0$. The key takeaway here is that having multiple ancestors which can result in the same or similar target sequence increases the likelihood of change. Certainly, the existence of several pathways to reach a particular outcome offers redundancy and adaptability. Yet, while "convergent" and "parallel" evolution are important in evolutionary theory (such as for demonstrating the predictability of evolutionary processes), a deeper analysis of them—which have been extensively discussed in the literature (Bolnick *et al.*, 2018; Cerca, 2023)—is outside the scope of our research.

## II.- Divergent Protein evolution

Under this model, the presence of a specific target sequence is not a prerequisite (see Figure 2**a**). The latter, rather than being a substantial disadvantage, presents an opportunity to discuss a model for a broader, albeit simplified, scenario of protein evolution. In this context, we will begin by examining the factors that influence evolutionary pathways and their effects, as they are crucial to the interpretation of a divergent evolutionary model.

### 1.- *The location and order of mutations*

The location and order of mutations may exert a substantial impact on the evolutionary pathways, such as turning a functional protein sequence into a nonfunctional or unfolded structural form; hence it is critical to understand their genesis. Whether a mutation is destabilizing or stabilizing is primarily determined by its location rather than the type of substituted amino acid (Vila, 2024a). Indeed, if the mutations happen in sites near or around the hydrophobic core of the native protein, there is a high chance of them being highly destabilizing; *e.g.*, it could speed up the folding rate by up to three orders of magnitude (Vila, 2024a). As a result, the location and order in which the mutations occur could be decisive for the protein evolution outcome. For instance, the effect of keeping the location but changing the order in which a mutation occurs is illustrated in Figure 2b through an oversimplified evolutionary scenario. If this were the case, the resulting total free-energy change (ΔΔG) could go above a certain threshold—around ±7.4 kcal/mol—beyond which the protein would denature or become non-functional (Vila, 2020). Finally, we must be aware that epistatic effects should be, by definition, sensitive to both the location and order in which mutations occur and, hence, could also alter the evolutionary pathway. However, because an accurate computation of epistasis is still uncertain (Vila, 2024b), such an effect is well beyond consideration.

## 2.- The post-translational modifications

Genomic mutations (Stella & Freed, 2024) and post-translational modifications (Ramazi & Zahiri, 2021), not just the former, should be considered the primary sources of evolution at the molecular level (Vila, 2024b). Particularly, post-translational modifications (PTMs) are not only important because of their enormous diversity—when compared to the 20 naturally occurring amino acids—but also because of their impact on proteins in terms of both structural and evolutionary points of view. However, the latter brings some degree of difficulty. For example, the presence of around 400 post-translational modifications (Ramazi & Zahiri, 2021) makes it difficult to determine, among other things, how they may influence the preferred pathways of protein evolution or mutational epistasis. Post-translational modifications will mainly affect solvent-exposed residues, whereas genetic mutations are not subject to such restrictions. It will have an important impact on protein evolution at the molecular level because it will assure us that PTMs are unlikely to lead to large free-energy protein stability changes against what could happen with amino acid substitutions, where such a guarantee does not exist. At this point, it is worth remembering that adding PTMs into naturally occurring amino acids as mutation sources does not change the upper bound limit of the available protein sequence space due to the strict constraints imposed by the folding rates (Vila, 2024b). These facts underscore the delicate balance between the potential for enhanced functional diversity from PTMs and epistasis effects, as well as the need for a thorough understanding of how all of this impacts on protein evolvability.

## 3.- The epistasis

Regarding epistasis effects, there are suggestions claiming that specific and nonspecific epistasis appear to affect the reversibility of evolution in different manners (Starr & Thornton, 2016). While examining the potential presence of a wide range of epistatic interactions is outside our scope of interest, it is worth mentioning that the solution to the epistasis problem, as mentioned in the Introduction Section, demands determining *how* the amino acid sequence encodes its folding—a problem equivalent to solving protein folding—which, as already noted, is a challenging undertaking (Vila, 2023a). Consequently, the determination of the molecular basis of mutational epistasis and their impact on both the repeatability and reversibility of protein evolution will remain unknown at least until the protein-folding problem is properly addressed (Vila, 2024b). This is the main reason *why* an accurate understanding of the epistasis impact at the molecular level is limited to a few specific applications (Weinreich *et al*., 2006; Sailer & Harms, 2017b). Nevertheless, a study of the molecular basis of epistasis is undoubtedly necessary if we are to understand *why* one evolutionary pathway is chosen over another (Weinreich *et al*., 2006).

### 4.- *Protein robustness*

The fact that proteins are robust to mutations (Vila, 2024a) reveals to us that they have evolved—since the beginning of life on earth (~$10^9$ years ago)—by reaching the most stable native states or preserving their marginal stabilities by limiting the free-energy changes due to mutations within a well-defined threshold (~7.4 kcal/mol, Vila, 2019) beyond which the protein becomes nonfunctional or unfolds (Martin & Vila, 2020; Vila, 2022). According to this viewpoint, the evolutionary process is presumably recurrent since most mutations should only marginally alter the native state stability of a protein. This perspective is significant as it suggests a connection between the predictability of protein evolution over time and the repeatability of the evolutionary process. Considering the significance of small protein marginal stability changes for protein evolvability and repeatability, it is pertinent to inquire: what proportion of single-point mutations (other than synonymous) results in a slight free-energy change in the marginal stability of naturally occurring proteins? Recently, we explored this problem by analyzing changes in the marginal stability of proteins ($\Delta\Delta G$), folding rates, and amide hydrogen-exchange protection factors (Vila, 2024a), with data from 341,860 mutations in which each of the naturally occurring amino acids was substituted on 17,093 sites of a set of 365 protein domains (Tsuboyama *et al*., 2023). The results indicate (after excluding synonymous mutations) that a significant fraction of mutations in naturally occurring proteins (~13%) lead to nearly neutral free-energy changes on the protein marginal stability—those for which $\Delta\Delta G$ changes are slightly stabilizing or destabilizing, *i.e.*, lower than ±0.1 kcal/mol (Matthews, 1995; Vila, 2024a). Note that this percentage is comparable to that leading to stabilizing mutations (~16%), although significantly lower than the destabilizing ones (~71%). These findings support the hypothesis that protein robustness, particularly in the face of destabilizing mutations, will make it possible for proteins to evolve preferentially while maintaining native state stability, increasing the likelihood of the process being repeatable and predictable.

Notably, the percentages given above represent the sensitivity of naturally occurring proteins to single-point amino acid substitutions, as around 87% of them will alter—mainly by worsening—their marginal stability beyond ±0.1 kcal/mol, regardless of the amino acid type substituted (Vila, 2024a). However, such a property of proteins has no relation to or impact on the mutation rate—equivalent to the rate of substitution of neutral mutations—which Kimura (1968), in a pioneering study, found to be relatively constant across species. Therefore, to better understand how genetic variation influences protein evolution, we need to know both how quickly mutations occur and how sensitive proteins are to those changes.

### 5.- How does protein sequence distribution impact on a divergent evolutionary model?

The first question to be answered is: How scattered could the protein sequence distribution be? Analyzing this problem may spark debate, as some scholars could argue that unconstrained evolutionary conditions will produce a vast ensemble of highly diverse protein sequences, while their native states are within the allowed range of free-energy variations (as shown in Figure 2b). Other scholars may claim that mechanisms exist to aid in maintaining protein sequence conservation in the face of evolutionary forces to change. For example, some protein species have virtually kept both structure and function since the beginning of aerobic life on Earth billions of years ago (Margoliash *et al.*, 1965). Consistent with this viewpoint and the Darwinian concept of evolution as descent with modification (Darwin, 1859), we have demonstrated that proteins should have only a fraction of their amino acid sequence prone to mutations (Vila, 2024a). We have arrived at this conclusion after demonstrating that the average mutation rate per amino acid ($\chi$) should be ~1.6 (Vila, 2024a). The condition $\chi < 2$ implies that only a fraction of a protein can tolerate mutations, regardless of their fold-class, sequence, or size; otherwise, the total number of possible functional protein sequences will surpass its allowed upper bound limit (~$10^{16}$ to ~$10^{19}$), a threshold imposed by thermodynamic hypothesis validity (Vila, 2024b). The existence of certain conserved, although non-functional, residues (in *c*-type cytochromes and globins) that play a crucial role during protein folding (Bychkova *et al.*, 2024) can be rationalized—from the above perspective—by acknowledging that not all residues of naturally occurring proteins are allowed to mutate. However, knowing which protein residues will not mutate requires significant time and effort (Bychkova *et al.*, 2024). While this topic is fascinating, it is outside the scope of our research.

Regardless of the solution to the above-posed dilemma, we can always identify—from a thermodynamic statistical point of view and independently of the total number of possible functional proteins—one sequence ($u$) that possesses the highest relative probability $\Omega$ ($P_u/P_{wt}$), although not the evolutionary pathways/trajectories leading to it. This uncertainty about forecasting evolutionary routes emphasizes the complexity and unpredictability of evolutionary processes, which are influenced by a wide range of factors, some of which have already been discussed here. Then, once the most probable sequence has been identified, the following evolutionary question arises: Can we anticipate what will happen next? All evidence suggests that the subsequent evolutionary steps are not envisioned to vary beyond the foreseen (as shown in Figure 2b), since most mutations that preserve biological function are expected to be neutral (Kimura, 1968) or nearly neutral (Ohta, 1973)—with respect to their fitness effects—whereas those that do affect fitness or introduce novel biological functions are likely to be predominantly destabilizing (Wang *et al.*, 2002; Tokuriki *et al.*, 2008; Jensen *et al.*, 2018). As a result, the assumption that proteins should evolve preferentially in the direction of preserving their native state marginal stabilities is neither incompatible with *retaining* biological functions or acquiring *new* ones—as a result of primarily destabilizing mutations—nor inconsistent with pieces of evidence indicating that no specific activity-stability tradeoffs are associated with the acquisition of new functions (Tokuriki *et al.*, 2008).

To conclude, the existence of an "effective" target sequence—for a divergent protein evolution model—enables us to assume, without losing generality, that *all* conclusions drawn from the analysis of convergent protein evolution (covered in Section I) will remain valid, making a divergent protein evolution model likely to be repeatable. For instance, there is evidence indicating that the evolution of hemoglobin or cytochrome *c* in various species can be traced back to a common ancestor. However, whether this showcases that protein evolution is repeatable, just a consequence of natural selection (Ryan, 2009), or the outcome of both remains to be proved.

## A molecular approach to protein evolution

So far, we have shown that proteins will evolve, in the long run, by adhering to the most probable sequence from a thermodynamic statistical perspective, regardless of the evolutionary model chosen for their analysis. The chosen simplified nature of the protein evolutionary process has enabled us to focus on the primary elements determining the probability of their recurrence, namely those impacting on the marginal stability of proteins. Nonetheless, it has left an important issue in structural and evolutionary biology unanswered . Specifically, what molecular mechanism would allow the proteins to evolve after a mutation while retaining their structure and function? To solve this issue, firstly, the thermodynamic hypothesis must be fulfilled at each evolutionary stage; otherwise, proteins will unfold or become nonfunctional. Secondly, consider the native state as a structure surrounded by an ensemble of high-energy native-like folds that coexist in fast-dynamic equilibrium (Lange *et al*., 2008; Ono *et al*., 2024)—a proposal in line with convincing theoretical simulations of the amide hydrogen exchange mechanism on proteins (Vendruscolo *et al*., 2003). Mutations or milieu alterations (such as pH, temperature, ionic strength, and so on) shake the stability of the native state of proteins, making possible the redistribution of the ensemble folded state ratio (Vila, 2022)—determined by its Boltzmann factors—which results in a new global free energy minimum that will now be represented by another (once high-energy) conformation of that ensemble. Within this perspective, it is possible to envision how proteins could evolve upon mutations and make the process repeatable while preserving their structure and function. This inference is consistent with a recent study (based on compiled data from 30 years of tracking morph frequencies across 10 replicate populations of a stick insect in the wild) that reports highly repeatable evolutionary fluctuations through time (Nosil *et al*., 2024).

## Conclusions

How protein topology, mutations, epistasis, recombination, natural selection, and genetic drift, among other factors—such as the impact of environmental pressures and epigenetic modifications—influence the pathways/trajectories of proteins through the sequence space in their search for functional forms is mostly unknown. Among all of these factors impacting protein evolvability, we have focused our efforts mainly on determining how protein evolution at the molecular level, specifically due to protein mutations, may impact either protein thermodynamic stability or evolutionary pathways/trajectories, as well as identifying the main factors that make protein evolution repeatable. In the pursuit of these goals, we explored convergent and divergent evolution models from a statistical-thermodynamics perspective, concluding that there is always an "effective" target sequence and, as a result, every protein evolutionary process at the molecular level could be evaluated as convergent evolution. This conclusion aligns with sound evidence that proteins have evolved since the beginning of life on Earth, roughly a billion years ago, in one direction: toward preserving their native state marginal stability—which is a direct consequence of the validity of the thermodynamic hypothesis—regardless of whether the evolutionary process is convergent or divergent. The analysis also leads us to the conclusion that protein evolution should be repeatable, as long as most mutations range from stabilization to slight destabilization of the native state, which is guaranteed by the thermodynamic hypothesis fulfillment, which provides a physical substrate for both the neutral or nearly neutral theory of evolution to occur (Martin & Vila, 2020) and the mutational robustness of proteins. Thus, proteins should evolve while preserving their structure and function.

As a whole, the demonstrated relationship between the thermodynamic hypothesis and protein evolution at the molecular level emphasizes the importance of fundamental physical principles in guiding evolutionary processes, and it provides a better understanding of how proteins can evolve through mutations and make the process repeatable while retaining their structure and function. Undoubtedly, the evolution of proteins is a complex process that is shaped by both biological and physical factors, some of which we have analyzed here.

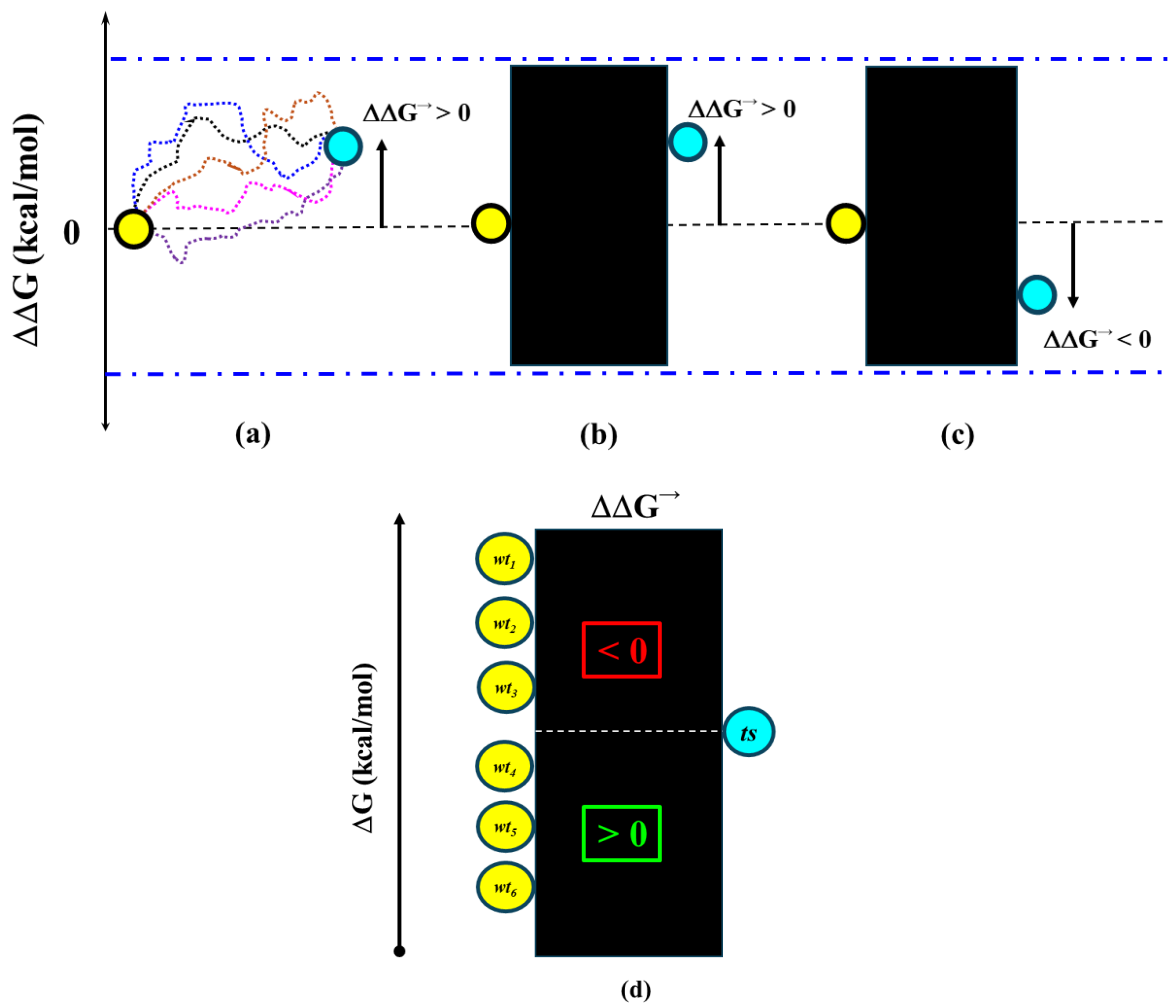

**Figure 1**. The cartoon illustrates possible scenarios—in terms of ΔΔG value—for a convergent evolutionary model. The horizontal blue dash-dot lines (in panels **a-c**) designate a threshold for ΔΔG (~ ±7.4 kcal/mol) beyond which a protein will unfold or become nonfunctional. Panel (**a**) illustrates an evolutionary process that is likely to be repeatable (ΔΔG > 0), albeit unlikely to be reversible. In this panel, every feasible evolutionary pathway (illustrated by different colored dashed lines) starts from a wild-type protein (yellow-filled circle) and ends up on a "target sequence" (depicted by cyan-filled circles). Panel (**b**) illustrates a reasonable alternative representation for panel (**a**), *i.e.*, as a black box, because, from a thermodynamic viewpoint, ΔΔG is a state function, and so the results are independent of evolutionary pathways. Panel (**c**) illustrates an evolutionary process that is unlikely to be repeatable (ΔΔG < 0), although likely to be reversible. Panel (**d**) illustrates—by using a black box representation of the evolutionary process—an example of multiple ($j$ = 6) ancestors (yellow-filled circles) converging to the same target sequence (cyan-filled circle). The arbitrarily chosen distribution for ΔG$_{wtx}$ (for $x$ = 1 to 6, corresponding to each of the six protein ancestors) shows that half of them lead to ΔΔG > 0, and hence a likely repeatable evolutionary process; whereas the remaining ones (with ΔΔG < 0) do not.

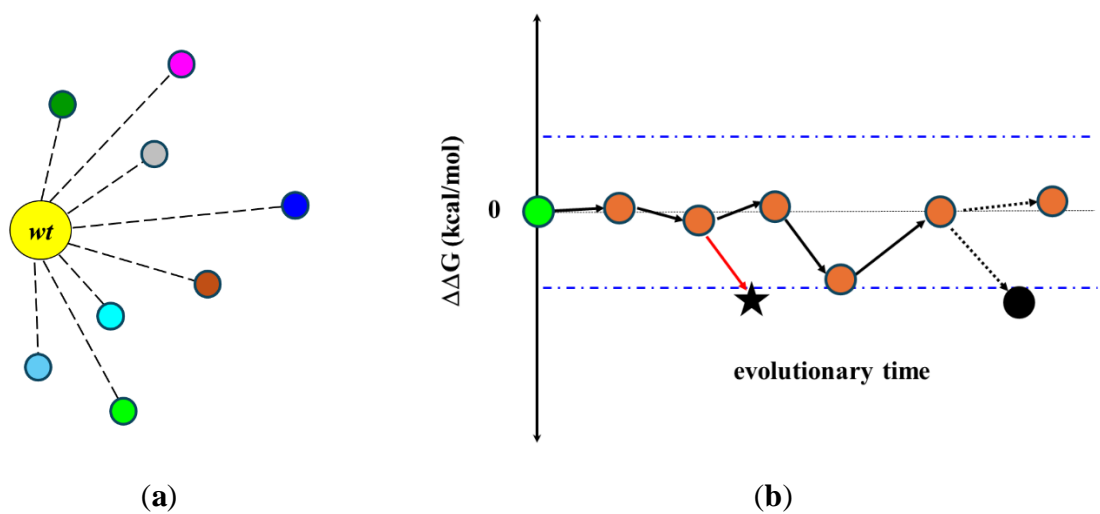

(**a**)                                                    (**b**)

**Figure 2**. Panel (**a**) illustrates a snapshot obtained after a certain timeframe of a divergent evolutionary model. In this panel, each ending evolutionary path in the protein sequence space is represented by a different-colored, filled circle. All eight pathways displayed—among thousands of other possible ones—start from the same wild-type (*wt*) protein (shown by a yellow-filled circle) and reach their final protein sequence through different numbers and types of mutation—such as amino acid substitutions or posttranslational modifications—each of which is indicated by a black-dash line. Panel (**b**) illustrates a potential evolutionary pathway, originating from the most probable protein sequence (arbitrarily) selected from panel (**a**). The arrows denote mutational steps, while the brown-filled circles indicate the $\Delta\Delta G$ value following each mutation. Such an evolutionary pathway—here illustrated for only six mutational steps—could end in either a functional or a nonfunctional protein sequence (highlighted as a black-filled circle). These possibilities are a consequence of the fact that any protein sequence showing a $\Delta\Delta G$ beyond a given threshold—indicated in the figure by the horizontal blue dash-dot lines ($\sim \pm 7.4$ kcal/mol)—will unfold or become nonfunctional. Panel (**b**) also illustrates what could happen if the order of mutations is altered. This scenario is presented after the second mutational step when the third (stabilizing) mutation is replaced by the fourth (destabilizing) mutation (highlighted here as a red arrow). As a result, the evolutionary pathways abruptly end with an unfolded or nonfunctional protein sequence (shown by a black-filled star).